\documentclass{aa}

\usepackage{natbib}
\bibpunct{(}{)}{;}{a}{}{,} 
\usepackage{graphicx}

\newcommand{\kss}{km~s$^{-1}$ }
\newcommand{\ks}{km~s$^{-1}$}
\newcommand{\cthead}[1]{\multicolumn{1}{c}{#1}}
\begin{document}

\title{Torsionally excited methanol at 44.9~GHz}

\author{M.A. Voronkov\inst{1} \and
        M.C. Austin\inst{2,3} \and
	A.M. Sobolev\inst{4}}

\offprints{M.A. Voronkov, \email{voronkov@tanatos.asc.rssi.ru}}

\institute{Astro Space Center, Profsouznaya st. 84/32, 117997
           Moscow, Russia \and
	   The Catholic University of America, 620 Michigan Ave.,
	   Washington, DC, 20064 USA \and
	   Maria Mitchell Observatory, 59 Milk st., Nantucket, MA, 02554 USA \and
	   Astronomical Observatory, Ural State University, Lenin st. 51,
           620083 Ekaterinburg, Russia}

\date{Received / Accepted }

\abstract{
Using the Haystack Observatory 37-m radio telescope we have
undertaken a search for emission in the $2_0-3_1$~E rotational transition of
methanol in its first torsionally excited state ($v_t$ = 1) at 44.9 GHz. We
examined seven galactic sources -- six strong emitters of Class~II methanol
maser lines and \object{Orion KL}, the only source where this line had been previously
detected. We confirm (at a level of $5\sigma$) the previous detection and
report two new detections -- a reliable ($9\sigma$) detection in \object{W3(OH)}
and a marginal ($3.5\sigma$) detection in \object{NGC 6334F}. Upper limits for other
sources are presented. Although we did not see obvious
signatures of maser amplification in this transition in any source, arguments
in favor of weak masing in \object{W3(OH)} are presented.
}

\maketitle

\section{Introduction}
Of the molecular species observed in
interstellar space, methanol has by far the
largest number of detected transitions, both in
thermal and maser emission. Interstellar
methanol masers are associated with regions of
active star formation, and fall into two
classes \citep{men91a}. Class~I masers are normally seen apart from compact
continuum sources, while Class~II masers are found close to ultra-compact
\ion{H}{ii} regions. The difference between the two classes is likely due to
a difference in pumping mechanism: collisional excitation
produces Class~I masers while radiative excitation produces
Class~II masers \citep[e.g.,][]{cra92}.
Torsionally excited interstellar methanol was first
detected by \citet{lov82}
who observed the $6_1-5_0$~E, $1_0-2_1$~E,
$2_1-1_1$~E, and $2_0-1_0$~E rotational transitions of the first
torsionally excited state ($v_t=1$) towards the
\object{Orion Kleinmann-Low} (KL) nebula. Later, other
transitions between torsionally excited levels were detected by \citet{men86b},
with the $10_1-11_2$~A$^+$, $v_t=1$ transition probably being inverted
in \object{W3(OH)}. \citet{sob94} suggested that pumping cycles through the
first two torsionally excited states may play an important role
in the pumping of the brightest Class~II methanol masers
at 6.7~GHz and 12~GHz.
Such a model is able to explain the observed brightnesses at 6.7~GHz and 12~GHz
as well as the ratio of their intensities. This model predicts
weak maser emission in the $2_0-3_1$ transition of the
first torsionally excited state of E-type methanol at 44.9~GHz \citep{sob97},
although no obvious correlation between 44.9~GHz brightness and that of
other known transitions was expected.
Thus, observations of this transition may be a valuable test of the Class~II
methanol maser pumping model.
The goal of this work is to search for $2_0-3_1$~E, $v_t=1$
emission towards some of the brightest known Class~II masers.
In addition, we reobserved this line in \object{Orion KL}, which is the only
source where 44.9~GHz emission was previously detected
\citep{sai89}.

\section{Observations}

For our observations in June and July of 2000 we used the
radome-enclosed 37-m telescope at the Haystack Observatory, Westford, MA.
A 35.5$-$49~GHz tunable maser amplifier receiver was used at the
$2_0-3_1$~E, $v_t=1$ rest frequency 44955.8$\pm$0.05~MHz
\citep{tsu95} with typical system noise temperature between 150~K and 400~K.
The notable exception is the low declination source \object{NGC 6334F}
for which the system temperature was often as high as 700~K.
All system temperatures are given on a corrected antenna temperature
($T_{\mathrm A}^*$) scale. The backend
was an autocorrelator operating with a bandwidth of 5.93~MHz split
into 8192 channels. During postprocessing, each of 32 adjacent channels were
averaged and then additional Hanning smoothing was applied,
yielding 128 channels with
a velocity resolution of 0.309~\ks. At 44.9~GHz, the beamwidth was 45~arcsec
and the aperture efficiency was about 0.27. A corrected antenna
temperature of 1~K corresponds to a flux density of 9.5~Jy.
From observations of Jupiter and Venus we estimate that
this calibration is accurate to about $\pm$30\%.
The observations were carried out in the beam switching mode with
a switching frequency of 10 Hz. The pointing accuracy was about 12\arcsec.

\section{Results and individual sources}

\begin{figure}
\resizebox{\hsize}{!}{\includegraphics{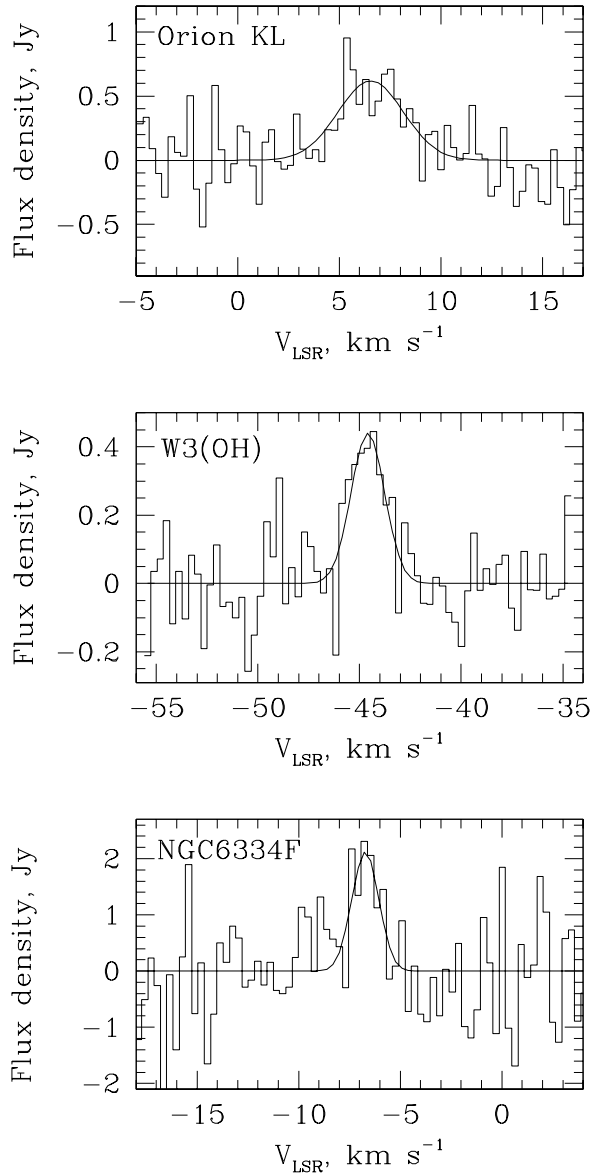}}
\caption{Spectra of the sources detected at 44.9~GHz. Solid line represents
Gaussian fit.}
\label{spectra}
\end{figure}

Our source list contains 7 sources, 6 known bright Class~II methanol
masers and \object{Orion KL}, which is a Class~I methanol maser.
In spite of the rather high noise level
we detected 44.9~GHz emission towards \object{W3(OH)}
(at a level of $9\sigma$)\footnote{The signal to noise ratio
was determined as the quotient of the integrated flux density and its
uncertainty, both given in the column~5 of Table~\ref{tab_res}.}
and \object{Orion KL}  ($5\sigma$).
For \object{NGC 6334} we present a marginal detection ($3.5\sigma$).
These three spectra are shown in Fig.~\ref{spectra}.
The other four sources brought no detections. Table~\ref{tab_res} contains
observed positions and velocities as well as achieved rms noise levels
for all observed sources. The table also shows results of Gaussian fitting
for the three sources with detected 44.9~GHz emission features.
Fitting errors are given in parenthesis. 

\begin{table*}
\caption{Sources observed at 44.9~GHz.}
\label{tab_res}
\begin{center}
\begin{tabular}{llllllll}
\cthead{Source} & \cthead{$\alpha_{1950}$} &\cthead{$\delta_{1950}$} &
\cthead{V$_{LSR}$}&\cthead{$\int S(v)\;dv$}&\cthead{$S_\nu$}&
\cthead{$\Delta v^a$}&\cthead{$1\sigma$-rms}\\
& \cthead{(h~~m~~s)} &\cthead{(\degr~~\arcmin~~\arcsec)} &
\cthead{(\ks)}&\cthead{(Jy~\ks)}&\cthead{(Jy)}&
\cthead{(\ks)}&\cthead{(Jy)}\\
\hline
\object{Orion KL} & 05 32 46.9& $-$05 24 23 & +6.5\hfill(0.4) & 2.5 (0.5) & 0.6 & 3.8 (1.0) & 0.3\\
\object{S252}     & 06 05 53.5& $+$21 39 02 & +11 &&&& 0.3\\
\object{W3(OH)}   & 02 23 16.5& $+$61 38 57 & $-$44.6~\hfill (0.1) & 0.9 (0.1) & 0.4 & 1.9 (0.3) & 0.1\\
\object{NGC 6334F} & 17 17 32.3& $-$35 44 04 & $-$6.7\hfill (0.2)& 3.5 (1.0) & 2.1 &  1.6 (0.5) &  0.9\\
\object{9.62$+$0.19}& 18 03 16.0& $-$20 31 53 & $-$0.7 &&&& 0.3\\
\object{W48}      & 18 59 13.1& $+$01 09 07 & $+$43 &&&& 0.2\\
\object{W75N}     & 20 36 50.4& $+$42 27 23 & $+$5  &&&& 0.3\\
\hline
\end{tabular}
\end{center}
\noindent 
~~~~~~~~~~~~~~~~
$^a$~$\Delta v$ is the full width at
half maximum (FWHM).
\end{table*}

\subsection{Orion KL}
As mentioned above, \object{Orion KL} is the only source for which
a detection of the
$2_0-3_1$,
$v_t=1$ line at 44.9~GHz was previously reported \citep{sai89}. Our values
for the linewidth
and the peak flux density are in agreement with those
observations within the errors of our measurement.
The peak radial
velocity of 6.5~\kss is typical for this region, e.g., emission of 
the hot core component (HC) peaks at velocities about $3-5$~km~s$^{-1}$
and emission of the methanol emission core (MEC, located about 10\arcsec~south
of our observed
position) peaks at $7-8$~km~s$^{-1}$~\citep{joh84, men86b}.

\citet{wil89} found the peak of the $10_1-9_2$~A$^-$ methanol line emission
at a velocity of about 8.0~\ks. They attributed the narrow component in the
spectrum of this line
to the MEC, and a broad and weak component to the HC.
It should be mentioned that the $10_1-9_2$~A$^-$ line is formed via a
transition between levels
with energies 140~K above the ground, which is rather low in comparison
to about 300~K
for the levels of the torsionally excited transition studied here.
\citet{min93} mapped the $15_3-14_4$~A$^-$ line emission of methanol,
originating from levels at 329~K above the ground. The map shows a
source of elongated shape between the HC and the MEC, and peaks
at the MEC, the radial velocity of the maximum being
7.5~\ks.
The same velocity and position properties were observed for
some torsionally excited lines seen towards \object{Orion KL}~\citep{men86b}.
So, most of the $2_0-3_1$, $v_t=1$
emission is probably associated with the MEC. 
The velocity
difference of about 1~km~s$^{-1}$ may be attributed to fitting errors
because the line may have a complex non-Gaussian profile similar to that
reported for other lines~\citep{min93,joh84}, as well as to the 50~kHz uncertainty
in the adopted rest frequency which corresponds to an uncertainty in velocity
of about 0.33~km~s$^{-1}$.
If the 44.9~GHz emission comes from the MEC the flux density shown in
Table~\ref{tab_res} is slightly underestimated (we would detect about 90\%
of the total flux) because the position of
the MEC is about $\frac 14$ of the beamwidth away from the beam center.
\citet{cra01} detected the $1_0-2_1$~E, $v_t=1$ line at 93.1~GHz which
belongs to the same $\mathrm{J}_0-(\mathrm{J}+1)_1$~E line series as the observed
44.9~GHz transition. The peak
velocity and the peak flux density were about 7.6~\kss and 3.5~Jy, respectively.

\subsection{\object{W3(OH)}}
Towards \object{W3(OH)} the $2_0-3_1$~E, $v_t=1$ line has a peak at
$-$44.6~\ks. Such a
value allows us to associate this emission with the
gas located in front of the ultra-compact \ion{H}{ii} region
in \object{W3(OH)} rather than with the nearby
($\sim7$\arcsec apart) warmer region \object{W3(H$_2$O)}~--
the cluster of water vapor masers. According to \citet{wil91} the
gas around \object{W3(OH)} is responsible for emission at
velocities greater than $-$45~\kss while \object{W3(H$_2$O)} shows a velocity
of about $-$48~\ks.
The $10_1-11_2$~A$^+$, $v_t=1$ and $12_2-11_1$~A$^-$, $v_t=1$ lines
detected by~\citet{men86b} peak at a velocity of $-$44.0~\kss which
also implies an association with \object{W3(OH)}. The velocity of about $-$44.6~\kss
falls in the range covered by masers in \object{W3(OH)}, though weak masers in this source
usually have the strongest spike at a velocity
of $-$43.2~\kss \citep[e.g.,][]{sut01}.
The observed linewidth of the 44.9~GHz line in \object{W3(OH)}
is about 1.9~\kss while broad and probably non-maser lines in this
source usually have
widths 2 or 3 times larger \citep[e.g.,][]{sly95,kal97}. However, torsionally excited lines
detected by~\citet{men86b} near $-$44.6~\kss
are as narrow as the 44.9~GHz line in our observations. This most
probably reflects the fact that lines formed by transitions
between levels with high excitation energy
originate from a rather compact region (for example, near the
exciting source) and, hence, the velocity dispersion is small in
comparison to the majority of lines with considerably lower excitation energy.

\subsection{\object{NGC 6334F}}
\object{NGC 6334F} is a marginal detection: the peak flux density
is only a factor of 2 higher than the rms noise in the spectrum
(see Table~\ref{tab_res}).
However,~\citet{men89} detected many non-maser
lines including the torsionally excited $10_1-11_2$~A$^+$, $v_t=1$ line
at a velocity of about $-$6~\ks. This gives a hint that this feature
can be real. The maser velocity of about $-$10.1~\kss is not present
in our spectrum, so the line, if any, probably has a  quasi-thermal
origin. Towards this source \citet{cra01}
detected $1_0-2_1$~E, $v_t=1$ emission at
93.1~GHz produced by a transition from the same
$\mathrm{J}_0-(\mathrm{J}+1)_1$~E series as the 44.9~GHz line.
The velocity was about $-8$~\ks, and the line was interpreted as
quasi-thermal too.

\section{Discussion}
\subsection{Is the 44.9~GHz emission in \object{W3(OH)} a low gain maser?}

The weak 44.9~GHz line may represent
either a thermal feature
or a weak maser.
It is possible to prove the low gain maser
nature of the observed weak feature in \object{W3(OH)} by an analysis
of the excitation
temperature of the $2_1-3_0$~E transition of the ground torsional state at
19.9~GHz. The latter transition is important for two reasons.
Firstly, rather strong maser emission was
detected at 19.9~GHz towards \object{W3(OH)} \citep{wil85}.
Secondly, each level of the
44.9~GHz transition is connected by a strong allowed radiative
transition with only one level of the 19.9~GHz transition and
vice versa, while other transitions with a change of torsional
quantum number are forbidden.
These facts provide the basis for the following estimates.

\begin{figure}
\resizebox{\hsize}{!}{\includegraphics{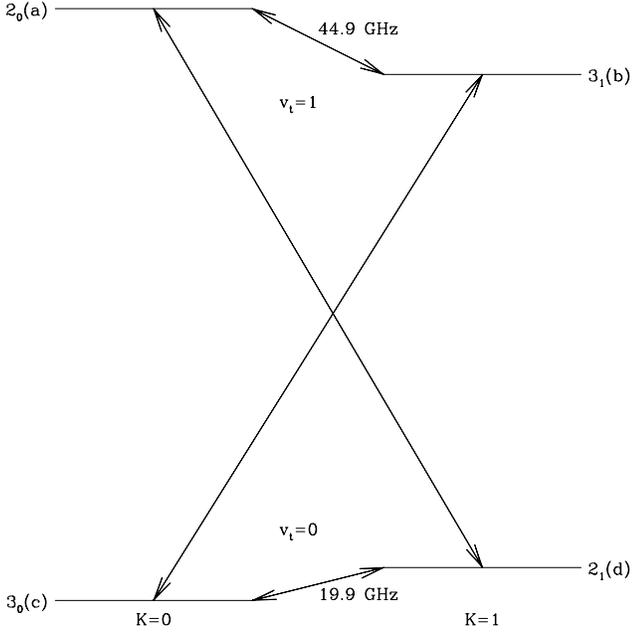}}
\caption{Levels of 44.9~GHz and 19.9~GHz transitions. Arrows
represent all allowed transitions (both upward and downward) between these
levels.}
\label{tex_rel.ps}
\end{figure}

The subsystem of the levels of the two transitions is given in
Fig.~\ref{tex_rel.ps}, where arrows show allowed transitions. The levels
$2_0$, $3_1$ of the first torsionally excited state and the levels
$3_0$, $2_1$ of the ground torsional state are labeled a,b,c,d respectively.
The formal definition of the excitation temperature corresponds to the
following relation:
\begin{equation}
\label{tex_relation}
\frac{\nu_\mathrm{ab}}{T_\mathrm{ex,ab}}=\frac{\nu_\mathrm{ad}}{T_\mathrm{ex,ad}}-
\frac{\nu_\mathrm{bc}}{T_\mathrm{ex,bc}}+\frac{\nu_\mathrm{dc}}{T_\mathrm{ex,dc}}\mbox{,}
\end{equation}
where $T_{\mathrm{ex,}ij}$~is the excitation temperature of the transition between
two levels, $i$ and $j$, and $\nu_{ij}$~is its frequency.
From  Fig.~\ref{tex_rel.ps} it is obvious that
\begin{equation}
\label{nu_relation}
\nu_\mathrm{ab}=\nu_\mathrm{ad}+\nu_\mathrm{dc}-\nu_\mathrm{bc}\mbox{.}
\end{equation}
Further, we are making the important assumption that both
transitions with a change of torsional quantum
number, $\mathrm{ad}$ and $\mathrm{bc}$,
have excitation temperatures equal to the temperature $T_\mathrm{R}$
of the external radiation field at the transition frequency
of about $6\times10^{12}$~Hz.

Methanol masers are generally assumed to be
pumped by diluted emission of the hot dust~\citep[e.g.,][]{sob94,sut01}.
Hence,
$T_\mathrm{R}$ can be derived from the dust temperature using guesses on dilution and
dust opacity. The best fit model of
\citet{sut01} yields $T_\mathrm{R}\approx83$~K while the brightness temperature at
the $\mathrm{ad}$ frequency is 0.3~K higher than that at the $\mathrm{bc}$ one.
In principle, such a difference helps inverting the 44.9~GHz transition (see considerations below).
However, for the sake of simplicity this difference will be neglected.

The equality of excitation and radiation temperatures is expected
when collisional rates for transitions between
different torsional states are negligible and is confirmed by modeling (Ostrovskii, priv. comm.). So, such equality
implies that for
transitions $\mathrm{ad}$ and $\mathrm{bc}$ (see Fig.~\ref{tex_rel.ps})
the value of $C/A$ is small compared to $B\overline{I}/A=W\tau_\mathrm{d}/(\exp\left(
h\nu/kT_\mathrm{d}\right)-1)$ and $C/A\ll1$. Here $A$ and $B$ are Einstein coefficients,
$C$ is the collisional transition rate and $\overline{I}$ represents
the average intensity at the transition frequency $\nu$. Further
parameters of the dust layer responsible for pumping are the temperature
$T_\mathrm{d}$, the dilution factor $W$ and
the optical depth $\tau_\mathrm{d}$. The experiments
conducted by \citet{lee74} in conjunction with a hard sphere approximation
yield C=$1.3\times10^{-4}$~s$^{-1}$ for methanol
collisions with para-H$_2$ at a hydrogen density of $10^6$~cm$^{-3}$
and a kinetic temperature of 100~K. The spontaneous emission
rates for both transitions are about 0.2~s$^{-1}$~\citep{mek99}, so,
$C/A\approx6.5\times10^{-4}$. Hence, for a dust temperature of 175~K, the
collisions are negligible in comparison to induced transitions if
$W\tau_\mathrm{d}\gg2.7\times10^{-3}$. The best fit model of \citet{sut01} has
$W\tau_\mathrm{d}=0.18$ ($W=0.5$ and $\tau_\mathrm{d}=0.36$ at a frequency around
$6\times10^{12}$~Hz). Note that collisional rates for transitions with a
change of torsional quantum number are believed to be considerably
lower than the values obtained using a hard sphere approximation
due to a large energy separation of the levels~
\citep{uma79,mas79}, and this increases the prevalence
of induced transitions.

It follows
from (\ref{tex_relation}) and (\ref{nu_relation}) that
\begin{equation}
\label{tex_final}
\frac 1{T_\mathrm{ex,ab}}=\alpha\left(\frac{1-\alpha}{\alpha T_\mathrm{R}}+\frac 1{T_\mathrm{ex,dc}}
\right)\mbox{,}
\end{equation}
where $\alpha=\nu_\mathrm{dc}/\nu_\mathrm{ab}>0$~--~the ratio of the frequencies.
The small difference in $\mathrm{ad}$ and $\mathrm{bc}$ excitation temperatures (see above)
results in an additional negative term $\Delta$ on the right side of (\ref{tex_final}),
increasing the degree of 44.9~GHz inversion
\begin{equation}
\label{delta}
\Delta=-\frac {\nu_{ad}}{\nu_{ab}T_\mathrm{R}}
\left(\frac{\Delta T_\mathrm{R}}
{T_\mathrm{R}+\Delta T_\mathrm{R}}\right)\approx\nu_{ad}\left(1-\alpha\right)
\frac d{d\nu}\;\frac 1{T_\mathrm{R}}\mbox{,}
\end{equation}
where $\Delta T_\mathrm{R}$ is the difference between
brightness temperatures of the radiation field at $\nu_{ad}$ and $\nu_{bc}$
(see Fig.~\ref{tex_rel.ps} for level description). When the pumping radiation
corresponds to black-body the correction equals to zero because the temperature of
the radiation is constant. In the considered case of diluted radiation of the
optically thin dust the value of $\Delta$ is
negative because $\alpha<1$ and the
radiation temperature grows with frequency.

For the case with $\alpha<1$
and $T_\mathrm{ex,dc}<0$ the $\mathrm{ab}$ transition will be
inverted if
\begin{equation}
\label{tex_cond}
-T_\mathrm{R}<\frac{1-\alpha}{\alpha}T_\mathrm{ex,dc}<0\mbox{.}
\end{equation}
This expression is directly applicable to the case of the 44.9~GHz line
in \object{W3(OH)} because $\alpha$ is about 0.44
and the 19.9~GHz transition is inverted.
So, to show that the 44.9~GHz line is inverted in \object{W3(OH)} one can
estimate the excitation temperature of the 19.9~GHz transition  and compare
it with the radiation temperature at frequencies
around $6\times10^{12}$~Hz.

The peak optical depth is determined 
by both the excitation temperature and the specific (divided by the
line width) column density of molecules in
the upper level of the considered transition. Assuming that the excitation temperature
is constant throughout the source and the
line profile is Gaussian, one obtains the following expression for the optical depth in 
the  19.9~GHz ($\mathrm{dc}$) line center:

\begin{equation}
\label{tau_final}
\tau_\mathrm{dc}=\frac{c^3A_\mathrm{dc}\sqrt{\ln 2}}{4\pi^{3/2}\nu_\mathrm{dc}^3}\left(exp(\frac{h\nu_\mathrm{dc}}
{kT_\mathrm{ex,dc}})-1\right)\left(\frac{N_\mathrm{d}}{\Delta V_\mathrm{dc}}\right)\mbox{,}
\end{equation}
where $A_\mathrm{dc}$~is the spontaneous decay rate and
$\Delta V_\mathrm{dc}$~is the full line
width at half maximum (FWHM) for the 19.9~GHz transition,
$N_\mathrm{d}$~is the column density
of the molecules at the level~$\mathrm{d}$. 
Transformation of (\ref{tau_final}) provides us with the following dependence of the 
excitation temperature on $\tau_\mathrm{dc}$ and the specific column density
$\left(N_\mathrm{d}/\Delta V_\mathrm{dc}\right)$
\begin{equation}
\label{tex_tau}
T_\mathrm{ex,dc}=\frac {h\nu_\mathrm{dc}}k\left[\ln\left(1-\beta\left(\frac{N_\mathrm{d}}{\Delta V_\mathrm{dc}}\right)^{-1}\right)
\right]^{-1}\mbox{,}
\end{equation}
where $\beta$ denotes the critical specific column density of the molecules in
the transition's upper level
\begin{equation}
\label{beta}
\beta=-\frac{4\pi^{3/2}\nu_\mathrm{dc}^3\tau_\mathrm{dc}}{c^3A_\mathrm{dc}\sqrt{\ln 2}}\mbox{.}
\end{equation}
When the specific column density is
equal to $\beta$ the
case of maximum possible inversion is realized, i.e., there are no molecules
in the lower level of the masing transition.

According to \citet{wil85} the 19.9~GHz maser in \object{W3(OH)} amplifies
the background emission by a factor of 100 at the peak velocity.
At the velocity of the 44.9~GHz emission the flux density is
about 5 times weaker,
yielding the optical depth $\tau_\mathrm{dc}\approx-3$. Adopting
the spontaneous decay rate $A_\mathrm{dc}=1.66\times10^{-8}$~s$^{-1}$~\citep{mek99}
and the rest frequency $\nu_\mathrm{dc}=19967.43$~MHz~\citep{xu97},
one obtains $\beta=1.4\times10^9$~cm$^{-3}$~s.

The dependence~(\ref{tex_tau}) is almost
linear for $\left(N_\mathrm{d}/\Delta V_\mathrm{dc}\right)\gg\beta$ and the
condition (\ref{tex_cond}) becomes
\begin{equation}
\label{final_cond}
\left(\frac{1-\alpha}\alpha\right)\left(\frac{h\nu_\mathrm{dc}}{k\beta}\right)
\left(\frac{N_\mathrm{d}}{\Delta V_\mathrm{dc}}\right)<T_\mathrm{R}\mbox{.}
\end{equation}
This yields $\left(N_\mathrm{d}/\Delta V_\mathrm{dc}\right)<9.2\times10^{10}$~cm$^{-3}$~s
for reasonable values of $T_\mathrm{R}\approx80$~K. The fraction of molecules populating
the $\mathrm{d}$ level with respect to the total number of molecules
(including both A and E species of methanol under the assumption of equal
abundances) is
estimated to be less than $3\times10^{-3}$ (equilibrium value for
a temperature of 175~K is
about $5\times10^{-4}$ and the value of the best fit model of~\citet{sut01} is about
$1.5\times10^{-4}$), giving the condition in terms of the full specific
column density $\left(N/\Delta V_\mathrm{dc}\right)<3\times10^{13}$~cm$^{-3}$~s.
\citet{wil85} report a linewidth $\Delta V_\mathrm{dc}<1$~\kss for the
19.9~GHz feature at a velocity near that of the 44.9~GHz line in \object{W3(OH)}.
Therefore the transition at 44.9~GHz is inverted if the line at 19.9~GHz
originates from a region with a methanol column density
$N<3\times10^{18}$~cm$^{-2}$.

Published estimates of the total column density of methanol ($N$)
in \object{W3(OH)} are well below this limit. Using the 96~GHz emission
line data 
\citet{kal97} estimated $N$ to be about $6.7\times10^{15}$~cm$^{-2}$.
Analysis of the 25~GHz absorption line data by \citet{men86a} provides 
$N\approx5.8\times10^{16}$~cm$^{-2}$ under
the assumption that all relevant excitation temperatures are equal to the
estimate of the rotational temperature $T_{rot}=73$~K.
The best fit model of
\citet{sut01} corresponds to $N\approx4\times10^{16}$~cm$^{-2}$.
So, the line detected at 44.9~GHz towards
\object{W3(OH)} is likely to be a low
gain maser. However, our method does not allow us to determine
the value of optical
depth and, hence, it cannot be used to predict 44.9~GHz line intensities
on the basis of observed masers at 19.9~GHz.
Further, the small difference in $\mathrm{ad}$ and $\mathrm{bc}$ excitation
temperatures occurring due to the difference in the brightness temperatures at
the respective transition frequencies (see above)
can provide an inversion of the 44.9~GHz even in the
case when the associated 19.9~GHz maser is weak or absent.
The gain difference as well as the difference in brightness of
amplified or pumping emission between maser spots is very influential.
Hence, the 44.9~GHz emission in \object{W3(OH)} may be associated with
a weaker 19.9~GHz
feature while the brightest  19.9~GHz feature has no 44.9~GHz
counterpart. 

The estimates of the 44.9~GHz optical depth using a single line
are very rough and arbitrary. However, the optical depth can be estimated
from the flux ratio under assumptions that the background amplification
dominates in the 44.9~GHz maser output and that the continuum and the line
sources in \object{W3(OH)} have the same diameter. The value of the line flux
$F_\mathrm{l}=0.4$~Jy is small compared to the continuum flux
$F_\mathrm{c}=3$~Jy~\citep{wil91}. So, the value of the optical depth is
about $\tau\approx-F_\mathrm{l}/F_\mathrm{c}\approx-0.13$.

\subsection{Non-detections of masers}
Except for \object{W3(OH)}, no reliable emission has been detected towards
other bright Class~II methanol masers. The marginal detection in \object{NGC 6334F}
is at a velocity near $-$6.7~\ks, which is different from the typical
maser velocity of $-$10.1~\kss and gives an argument in favor of its
non-maser nature.
It was shown that the inversion of the 44.9~GHz transition is expected,
if the transition at 19.9~GHz  is masing.
Strong 19.9~GHz masers have been detected both in \object{W3(OH)}~\citep{wil85} and
\object{NGC 6334F}~\citep{men89}. However, in our spectrum there is not any hint of
44.9~GHz emission at the velocity of the 19.9~GHz maser in \object{NGC 6334F}.
This can be explained by poor sensitivity of our \object{NGC 6334F} data.
Other possibilities are the
smaller (by absolute value) optical depth and the smaller
brightness of background emission in comparison to \object{W3(OH)}.

\subsection{Application to other lines}
In the case of a rather strong maser in the ground state transition 
the method described above can be used to predict the inversion of
transitions between the levels of the first torsionally excited state
with the same quantum numbers K (see Fig.~\ref{tex_rel.ps}) and
close quantum numbers~J ($\Delta\mathrm{J}=0,\pm1$). Therefore, we can
consider the $1_0-2_1$~E, $v_t=1$ transition at
93.1~GHz as $\mathrm{ab}$ instead of the
$2_0-3_1$~E, $v_t=1$ transition at 44.9~GHz.
In this case $\alpha\approx0.21$ and the relation~(\ref{final_cond})
requires a methanol column density $N<10^{18}$~cm$^{-2}$ on the
93.1~GHz line inversion, which is about half an order of magnitude less then
that for the 44.9~GHz line. However, the estimates of the column density cited
above satisfy the relation as well.
Therefore, the transition at 93.1~GHz seems also to be inverted
in W3(OH). As was mentioned already, the inversion of the line can not
be considered as an indication of observable masers, because the actual
gain and, hence, observed
brightness may be arbitrary. Information on the 93.1~GHz line in
W3(OH) is not published at present.

The strongest methanol masers were found in the $5_1-6_0$~A$^+$ transition at
6.7~GHz~\citep{men91b} and the $2_0-3_{-1}$~E transition at
12.2~GHz~\citep{bat87}. However, corresponding
lines of the first torsionally excited state have very high
frequencies (about $10^{12}$~Hz) and very low $\alpha$. So, the described
method is not applicable. The situation
for the bulk of known Class~II maser transitions is similar, both
for the A and E species of methanol. 

Important exceptions, besides the $2_1-3_0$~E transition at
19.9~GHz, are the masers in the $9_2-10_1$~A$^+$ transition at 23.1~GHz and
the $8_2-9_1$~A$^-$ transition at 28.9~GHz detected by \citet{wil84,wil93}.
Corresponding transitions in the first torsional state are
$8_1-9_2$~A$^+$ at 118~GHz, $9_1-10_2$~A$^+$ at 69.6~GHz, $10_1-11_2$~A$^+$
at 20.9~GHz, $7_1-8_2$~A$^-$ at 169~GHz, $8_1-9_2$~A$^-$ at 121~GHz and
$9_1-10_2$~A$^-$ at 73.8~GHz. However, ground state masers at 23.1~GHz and
28.9~GHz are usually weaker than the 19.9~GHz maser
and for most of the mentioned torsional transitions the value of
$\alpha$ is smaller than that in the case of the 44.9~GHz line.
So, the tendency of these torsional transitions to be inverted in the
sources with maser emission at 23.1~GHz or 28.9~GHz is less pronounced.

Regarding
the $10_1-11_2$~A$^+$, $v_t=1$ transition at 20.9~GHz,
\citet{men86b} observed this transition and concluded that it is probably
inverted in W3(OH). The method described
here cannot be directly applied
to this transition because $\alpha$ is greater than 1.

\subsection{Thermal nature of 44.9~GHz emission in Orion KL}
Orion KL is a Class~I methanol maser source and displays no
maser emission at 19.9~GHz~\citep{wil85} as is common for this class
of objects. We cannot draw any conclusion on
44.9~GHz inversion in this case.
\citet{men86b} performed
a Boltzmann analysis using a rotational diagram method~\citep{joh84}
for a number of methanol lines observed towards
Orion KL including torsional excited lines.
In general, the
data were consistent with the assumption of LTE, yielding a rotation
temperature of 139~K and a methanol column density of about
$10^{17}$~cm$^{-2}$ under the assumption that
the source size is 30\arcsec.
Our observations yield the integrated flux density
$\int S(v)\;dv=2.5$~Jy~km~s$^{-1}$ for Orion KL (see Table~\ref{tab_res}).
Assuming $A=3.78\times10^{-7}$~s$^{-1}$~\citep{mek99} for the 44.9~GHz
transition, one
obtains
$\mathrm{ln}\left(N_u/g_u\right)=29.0$, with a
source size of 30\arcsec. In the latter
equation $N_u$ is the column density of the molecule in the upper
level of the 44.9~GHz transition, and $g_u$ is the
statistical weight. The energy corresponding to the 44.9~GHz levels is
about 300~K above the ground level and our data are in
perfect agreement with the rotational diagram
by \citet{men86b}. The $1_0-2_1$~E, $v_t=1$ line at
93.1~GHz detected by~\citet{cra01} belongs to the same
$\mathrm{J}_0-(\mathrm{J}+1)_1$~E
line series as the observed 44.9~GHz transition and should
show similar behavior. \citet{cra01} find
$\int S(v)\;dv=11.8$~Jy~km~s$^{-1}$ for Orion KL. Along with
$A=3.74\times10^{-6}$~s$^{-1}$~\citep{mek99} and $E_u\approx291$~K
for the 93.1~GHz transition this
gives $\mathrm{ln}\left(N_u/g_u\right)=28.8$ which is in agreement with
the results of \citet{men86b}.
Hence, both torsionally excited lines at 44.9~GHz and 93.1~GHz have a
quasi-thermal nature in Orion KL.

\section{Summary}
We have observed the $2_0-3_1$~E, $v_t=1$ methanol line at 44.9~GHz towards six
bright Class~II methanol masers and Orion KL where this line was
previously observed. We confirm (at a level of $5\sigma$) the previous
detection and
report two new detections -- a reliable ($9\sigma$) detection in \object{W3(OH)}
and a marginal ($3.5\sigma$) detection in \object{NGC 6334F}. We show that
this line may be a low gain maser in W3(OH) and probably has a quasi-thermal nature
in \object{NGC 6334F}. The line in Orion KL has a quasi-thermal nature.

\begin{acknowledgements}
We would like to thank  
V.S.~Strelnitskii, V.I.~Slysh, S.V.~Kalenskii, and D.M.~Cragg for several
valuable remarks and helpful discussions which undoubtedly improved
the publication and A.B. Ostrovskii for the help with model data.
We also thank an anonymous referee for helpful
remarks.
This project was partially supported by the NSF/REU grant AST-9820555.
MAV and AMS were supported by the INTAS grant no. 97-1451.
MAV was also supported by the RFBR grants no. 98-02-16916
and no. 01-02-16902
and the Radio Astronomy Research and Education Center (project no. 315).
\end{acknowledgements}

\end{document}